\documentclass[a4paper, twocolumn]{article}

\usepackage{pifont}
\usepackage{hyperref}
\usepackage{graphicx}
\usepackage{fancyhdr}
\usepackage{amsmath}
\usepackage{lmodern} 
\usepackage{longtable} 
\usepackage{amssymb}
\usepackage{enumitem}
\usepackage{subfigure}
\usepackage{tabularx}
\usepackage{colortbl}
\usepackage{booktabs}
\usepackage{array}
\usepackage{multirow}
\usepackage{url}
\usepackage{xcolor}
\usepackage{float}
\usepackage{abstract}
\usepackage{flushend}

\newcolumntype{D}{>{\centering\arraybackslash}X} 

\begin{document}

\title{
  \bf Bridging Human Cognition and AI: A Framework for Explainable Decision-Making Systems.
}

\author{
  N. Jean, G. Le Pera\\
  \includegraphics[width = 40mm]{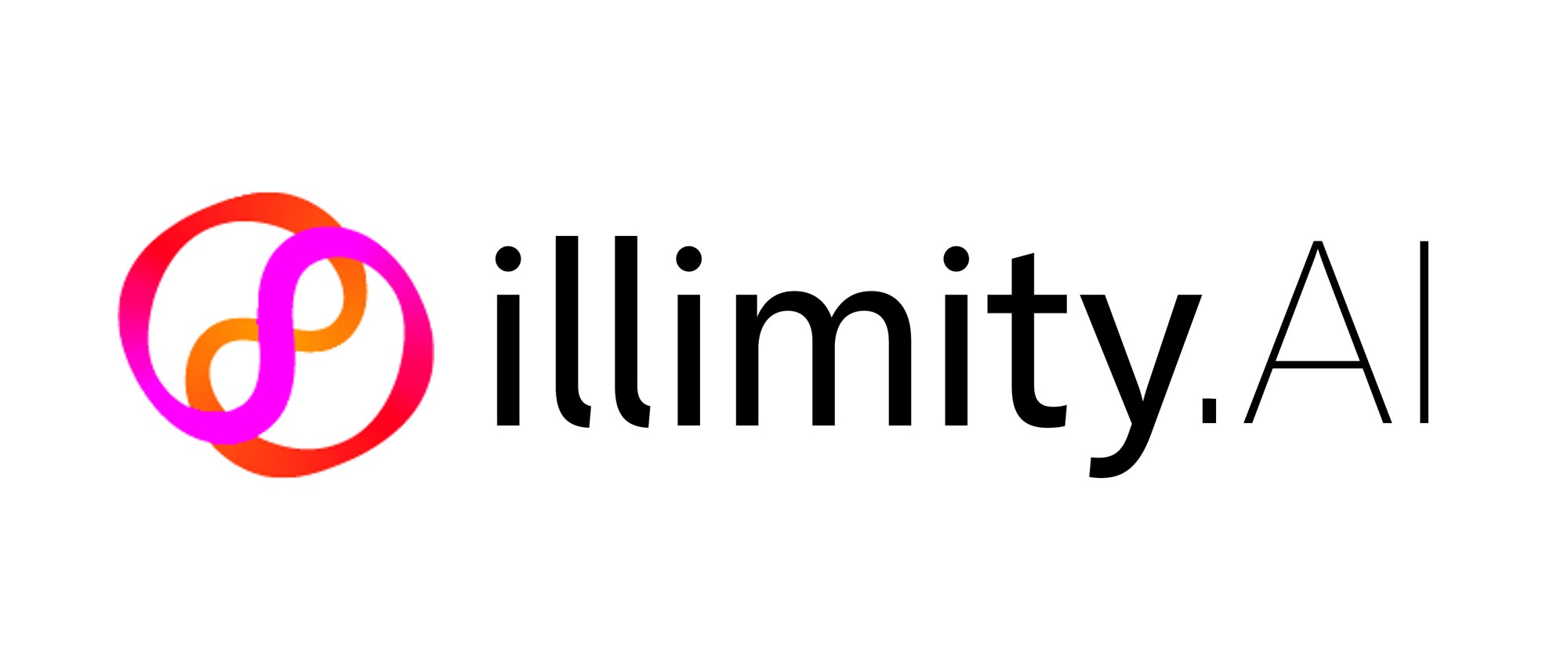}
}

\date{\today \\  [2em]
{\tt  Working paper\footnote{This paper reflects the authors' opinions and not necessarily those of their employers.}}
}

\twocolumn[
\maketitle

\begin{onecolabstract}
Explainability in AI/ML models is critical for fostering trust, ensuring accountability, and enabling informed decision-making in high-stakes domains. Yet, this objective is often unmet in practice. This paper proposes a general-purpose framework that bridges state-of-the-art explainability techniques with Malle’s~\cite{malle2004} five-category model of behavior explanation—\emph{Knowledge Structures}, \emph{Simulation/Projection}, \emph{Covariation}, \emph{Direct Recall}, and \emph{Rationalization}. The framework is designed to be applicable across AI-assisted decision-making systems, with the goal of enhancing transparency, interpretability, and user trust. We demonstrate its practical relevance through real-world case studies, including credit risk assessment and regulatory analysis powered by large language models (LLMs). By aligning technical explanations with human cognitive mechanisms, the framework lays the groundwork for more comprehensible, responsible, and ethical AI systems.
\medskip
\end{onecolabstract}

\small{\bf JEL} Classification codes: C61, C63, D83, G21\\
{\bf AMS} Classification codes: 91G40, 91G60, 68T01, 91E45\\
\bigskip
{\bf Keywords:} AI-assisted decision-making, explainability, behavior explanation, credit risk, regulatory technology, human-AI interaction
\bigskip
]
\saythanks

\section{Introduction}\label{sec:intro}

Explainability in AI/ML models is essential for building trust, ensuring accountability, and supporting effective decision-making in high-stakes domains. Leading technology companies—such as IBM~\cite{ibm_ai}, Microsoft~\cite{ms_responsible_ai}, and Google~\cite{google_explainability}—have issued broad guidelines to tackle explainability challenges in modern AI systems. However, developers often interpret these needs purely in technical terms, relying heavily on tools such as SHAP, LIME, and counterfactual analysis. These techniques, while valuable, typically neglect the cognitive frameworks humans use to interpret complex decisions. As a result, users frequently distrust the models, dismissing them as opaque black boxes—leading to outright rejection.

To address this disconnect, model developers and product managers need a structured framework to select appropriate explanatory tools that are aligned with users’ expertise, context, and expectations. The absence of such a framework often results in a reliance on shallow technical explanations—such as feature attributions or statistical summaries—that fail to satisfy real explainability requirements.

Explainability is too often conflated with mere algorithmic transparency. Tools like SHAP, LIME, and feature importance plots are effective at showing \emph{how} a model arrives at a decision, but they fall short of addressing \emph{why} that decision makes sense to the user in a given context~\cite{guidotti2018, gift2024, pelkmann2024}. This shortfall stems from two common misconceptions:

\begin{itemize}[label=\textbullet]
    \item \textbf{Explainability Equals Transparency}: A common belief is that exposing the inner workings of a model inherently leads to user understanding. However, presenting technical insights such as statistical thresholds or feature contributions often overwhelms non-technical users, leaving them without actionable clarity \cite{zhang2020, chen2022}.

    \item \textbf{One-Size-Fits-All Explanations}: Many AI/ML systems adopt a uniform approach to explainability, ignoring the diverse perspectives of their stakeholders—data scientists, business leaders, regulators, and end-users. Misaligned explanations can create confusion and mistrust, particularly when users expect different levels of detail or contextual relevance \cite{samek2019, schmitt2024}.
\end{itemize}

Such gaps highlight a crucial oversight: explainability is not solely about exposing the mechanics of an AI/ML model but about ensuring that explanations align with human cognitive processes \cite{malle2004}. To foster trust and accountability, explanations must resonate with the user's mental model, providing clarity and actionable insights tailored to their context \cite{tyagi2022}.
When users call models black boxes, they aren't asking for transparency—they want no box at all.

This paper introduces a novel framework for explainability that bridges technical transparency with human cognitive principles, breaking the perceived model's box. The following sections outline this framework and demonstrate its application in scenarios such as credit risk assessment and regulatory analysis with LLM agents, paving the way for more effective and human-centric AI/ML explainability \cite{guidotti2018_arxiv, ms_responsible_ai}.

\section{Theoretical Framework}\label{sec:literature}

To build the framework, we begin by outlining the underlying sociological foundations. Once these components are established, we proceed to the technical implementations.

\subsection{Emulation vs. Discovery Tasks (Observable vs. Unobservable)}\label{subsec:nace}

We begin our exploration of the framework from a sociological perspective. Malle’s theory introduces a fundamental distinction between \textit{observable} and \textit{unobservable} behaviors. While this distinction is intuitive in social contexts, it is more challenging to formalize within mathematical or modeling frameworks. To support both conceptual understanding and practical application, we adopt a parallel distinction introduced by Chen~\cite{chen2022}: \textit{Emulation} versus \textit{Discovery}. This reframing enables us to connect Malle’s categories to a well-known distinction in data science—namely, the one between supervised and unsupervised learning—while with some caveats preserving the essence of Malle’s original theoretical separation.

The distinction between emulation and discovery tasks is critical for managing explainability in AI systems. Chen~\cite{chen2022} introduces this dichotomy to analyze how explainability improves human understanding and, consequently, enhances decision-making. Specifically, explainability helps uncover the decision boundary, offering insights into the phenomenon partially captured by a model.

Unlike Chen~\cite{chen2022}, whose emphasis is on optimizing outcomes in automated decision-making (ADM), our primary focus is on fostering trust. Nonetheless, we rely on the same conceptual framework to ground our analysis.

First let us connect Chen's categories to practical distinction in terms of model setup.
In emulation tasks, users aim to replicate known behaviors with high precision, while discovery tasks involve exploring unknown outcomes, often under conditions of uncertainty. This distinction mirrors the difference between supervised learning—where models are trained on labeled data—and unsupervised learning—where no explicit target is provided. The presence of a target is closely tied to the user’s ability to emulate: users may have a clear expectation of the model’s outcome (emulation) or no prior expectation at all (discovery).

However, this distinction alone is not sufficient. Within supervised learning, we must further differentiate between cases where users are capable of evaluating the model’s output based on prior experience, and those where such judgment is not possible. When users can meaningfully assess a model’s predictions, they are effectively engaged in emulation; otherwise, the task leans toward discovery, as the correctness of the output cannot be easily determined.

Importantly, the boundary between emulation and discovery is not fixed—it depends on the user’s expertise. A task that represents emulation for an expert may still be a discovery process for a novice. Thus, the user's background plays a critical role in shaping the mode of explanation required for effective understanding and trust.

\begin{itemize}
    \item \textbf{Experienced users} can evaluate the model’s output based on their prior knowledge, making emulation tasks straightforward.
    \item \textbf{Novice users}, on the other hand, often engage in \textbf{discovery tasks}, where they lack a reference for correctness and must explore patterns or insights without predefined expectations.
\end{itemize}

In summary, emulation refers to a type of supervised learning in which the correct output is well-defined and can be readily assessed by humans. This allows users to verify whether the model’s predictions are accurate. However, this ability to judge correctness depends heavily on the user’s expertise. When such assessment is not feasible—due to uncertainty, novelty, or lack of domain knowledge—the task shifts to discovery.

This distinction between \textbf{emulation} and \textbf{discovery} is therefore not absolute—it varies based on the expertise of the users interacting with the model.

Tasks like sentiment analysis or NLP data labeling fall under emulation, as they involve replicable outcomes with clear decision boundaries. Discovery tasks, however, whether supervised or unsupervised, deal with scenarios where such clear discrimination is not possible. Credit scoring for loan approval and default probability forecasting are examples of discovery tasks, where uncertainty is high, and the user may lack a definitive ground truth or expertise for comparison.

Having clarified how to associate models with emulation and discovery tasks, we can now connect this distinction to Malle’s foundational concept of \textit{observable} versus \textit{unobservable} behaviors. The connection is intuitive: emulation tasks involve behaviors that can be judged and therefore must be observable. In contrast, discovery tasks—often dealing with unknown or uncategorizable outcomes—fall under the category of unobservable behaviors, as these outcomes cannot be readily judged.

\begin{center}
    \textit{Emulation} $\sim$ \textit{Judgeable target} $\sim$ \textit{Observable}  \\
    \textit{Discovery} $\sim$ \textit{Unjudgeable target} $\sim$ \textit{Unobservable}
\end{center}

By leveraging the concept of \textit{judgability} from the user's perspective, we can directly connect the technical properties of a model’s training target with the social categories outlined in Malle’s framework. This linkage bridges the gap between sociological theory and data science practice, making the framework applicable for building trust in ML/AI-based products.

We now turn to a deeper exploration of Malle's sociological theory and its alignment with a more technical perspective, focusing on the practical implementation of explainability methods. By following this sociological pathway, we demonstrate how theoretical distinctions can be translated into concrete mathematical models and techniques, ultimately enabling the delivery of effective explanations that enhance user trust.

The next critical distinction within Malle’s framework is that between \textit{intentional} and \textit{unintentional} behavior.

\subsection{Intentional vs. Unintentional (Folk Classification of Behavioral Events)}\label{subsec:embeddingtech}

In traditional product development, mathematical models are seen as tools for generating numbers, with machines functioning as components in a human-led process. However, as we scrutinize these models, they acquire a social dimension, with terms like "hallucinated," "creative," or "deceitful" often used to describe models such as large language models. This shift reframes "trust" as "social trust," requiring machines to navigate complex social interactions.

Malle's framework, as presented in \textit{How the Mind Explains Behavior}~\cite{malle2004}, highlights the importance of distinguishing not only between \textbf{observable} and \textbf{unobservable} behavioral events but also between \textbf{intentional} and \textbf{unintentional} actions. This distinction introduces a fundamental social dimension to the interpretation of behavior.

When interacting with a model, users may perceive the presence of an underlying intent, making explainability not merely a logical exercise but a deeply social issue. Trust and emotions play a crucial role, often outweighing purely rational explanations based on logical patterns.

The combination of Observable/Unobservable and Intentional/Unintentional categories results in four distinct groups, as shown in Table ~\ref{tab:explanation_types}. We've introduced the additional labels Systematic and Idiosyncratic, as they provide a clearer technical perspective compared to a simple Observable/Unobservable division.

\begin{table*}[h]
\centering
\renewcommand{\arraystretch}{1.2}
\setlength{\tabcolsep}{6pt} 
\footnotesize 
\begin{tabular}{|>{\centering\arraybackslash}m{4cm}|>{\centering\arraybackslash}m{5cm}|>{\centering\arraybackslash}m{5cm}|}
\hline
 & \textbf{Intentional (systematic)} & \textbf{Unintentional (idiosyncratic)} \\
\hline
\textbf{Observable (emulation)} & Actions & Behaviors \\
\hline
\textbf{Unobservable (discovery)} & Intentional thoughts & Experiences \\
\hline
\end{tabular}
\caption{Categorization of explanation types based on observability and intentionality.}
\label{tab:explanation_types}
\end{table*}

More in details the categories represent:

\begin{itemize}
    \item \textbf{Actions}: Systematic outcomes that users can replicate, where the system either reinforces or challenges beliefs.
    \item \textbf{Behaviors}: Idiosyncratic errors in emulation tasks, where model errors are difficult for users to understand.
    \item \textbf{Intentional Thoughts}: Users interact with the model to explore unknown realms, trusting the machine's deterministic path.
    \item \textbf{Experiences}: Idiosyncratic exploration where confidence is uncertain, and the machine operates unpredictably.
\end{itemize}

Once again, Malle helps us narrow down the relevant categories we need to consider.
Phenomenologically, he demonstrates that the most common cases involve \textit{Actions} and \textit{Experiences}, leading to the following association rule:

\begin{center}
     Emulation $\sim$ Observable $\rightarrow$ Intentional $\sim$ Actions \\
     Discovery $\sim$ Unobservable $\rightarrow$ Unintentional $\sim$ Experiences
\end{center}

Once a specific model type is identified, its explainability can be mapped to the most relevant category in Malle’s framework—namely, \textit{Actions} or \textit{Experiences}. The remaining two categories in Malle’s model, while theoretically important, are less frequently encountered from a phenomenological standpoint and can often be set aside in practical implementations.

This represents a key advantage of grounding explainability efforts in a sociological theory: it allows us to narrow the design space to what actually matters in practice. Instead of attempting to cover all possible explanatory categories—which would lead to unnecessarily complex or overly generic systems—we can focus on the two categories that most directly affect users’ trust and understanding.

When designing models and the products built around them, our primary concern should therefore be with \textit{Actions} and \textit{Experiences}. This sociologically-informed distinction provides a concrete, actionable foundation for selecting and implementing explainability tools that are well-aligned with the user's expectations and the system’s purpose.

\subsection{Explanation Modes in the Context of Actions vs. Experiences}\label{subsec:dimensionalityliter}

Now that we have successfully mapped the model's target traits to a specific user's perception, we can focus on the socially available methods required to express explainability.
As with any logical process, building a taxonomy is the first step in aligning explainability techniques with users' needs.

The explanation modes are conventionally classified as follows:

\begin{enumerate}[left=0pt, label=\textbf{\arabic*.}, itemsep=5pt, wide=0pt]
    \item \textbf{Reason Explanations (Actions)}: These involve revealing the underlying motivation or purpose behind an action, such as explaining someone's decision to work overtime as a commitment to meet a project deadline.

    \item \textbf{Causal History (Actions)}: These focus on narrating the sequence of events or factors leading to an outcome, like describing the circumstances that resulted in a business project's success or failure.

    \item \textbf{Enabling Factors (Actions)}: These identify conditions or elements that facilitated a specific behavior, such as explaining how effective communication enabled the successful completion of a team project.

    \item \textbf{Various Causes (Experiences)}: These refer to factors contributing to actions without conscious intention, such as:
    \begin{enumerate}[label*=\arabic*., left=15pt, itemsep=3pt]
        \item \textbf{Person Causes}: These attribute behavior to individual traits, like explaining someone's choice to pursue a challenging task due to their ambitious nature.

        \item \textbf{Situation Causes}: These link behavior to external circumstances, such as describing a person's decision to leave a party early because they felt unwell.

        \item \textbf{Various Interactions}: These emphasize the interplay between personal and situational factors, like explaining a team's success as the result of individual skills, teamwork, and organizational support.
    \end{enumerate}
\end{enumerate}

\vspace{1em}
All the previous types map naturally to various specific modes. Again social studies come to the rescue helping us in narrowing down the choices to the most commonly used in human environments \cite{malle2004}

Table \ref{tab:explanation_modes} extracted from \cite{malle2004} shows the most common cases with "xx", while a decreasing number of "x" marks a lesser usage.

\begin{table*}[h]
\centering
\renewcommand{\arraystretch}{1.3}
\setlength{\tabcolsep}{4pt} 
\footnotesize 
\begin{tabular}{|p{2.7cm}|p{2.3cm}|p{2.3cm}|p{2.3cm}|p{2.3cm}|p{2.3cm}|}
\hline
\textbf{Explanation/ Mode} &
\textbf{Knowledge Structures (HOW?)} &
\textbf{Simulation/ Projection (WHY?)} &
\textbf{Covariation (WHAT CAUSES IT?)} &
\textbf{Direct Recall (HISTORICAL SIMILARITY)} &
\textbf{Rationalization (LOGICAL REASON)} \\
\hline
\textbf{Cause (unintentional behavior)} & x & ( x ) & ( x ) & & \\
\hline
\textbf{Causal History} & xx & x & x & & \\
\hline
\textbf{Enabling Factor} & xx & & x & & \\
\hline
\textbf{Reason/Actors} & ( x ) & & & xx & x \\
\hline
\textbf{Reason/Observers} & xx & xx & ( x ) & & \\
\hline
\end{tabular}
\caption{Explanation Modes and Their Associations}
\label{tab:explanation_modes}
\end{table*}

This table is central to our work, as we will use it to select the most commonly used explanation modes. Summarizing, the modeller's task is now to determine whether the model is used for emulation or discovery. Once this is established, he will map it directly to either the Action or Experience category and implicitly to Actions vs Experiences. From there he will select one or more of the explanation modes listed in Table~\ref{tab:explanation_modes}. These are the tools we will use to break the black box conundrum of modern ML/AI models.

The selection of the correct explanation mode is a complex task. Malle suggests calibrating a series of equations against user feedback to choose the most appropriate mode, with his proposed framework concluding with a few caveats and examples.

The choice of mode depends on several factors, such as the difficulty of reproducing the behavior, the type of question we need to answer, and the availability of information, among others. In \cite{malle2004}, some general rules of thumb are proposed, which serve as an initial guide to understand what methods should be applied for specific use cases. We will rely on those hints in the next section to illustrate how to appply the framewrok to two practical examples, leaving a more numerical and adta driven analysis to further studies.

The framework will also be used to highlight which models should not be implemented. For this specific use no calibration on user's preferences, skills or data properties is needed.

\section{Practical implementation }\label{sec:methodology}

In the architecture specified for our model store, we have designed a framework that will soon be implemented to expose a set of explainability models, grounded in two core components: a model layer—extending the Alibi library~\cite{alibi}, a widely adopted explainability toolkit—and a data layer based on a vector database such as Qdrant~\cite{qdrant}.

Each model intended for use in an Automated Decision-Making (ADM) context will be required to support this standardized interface. In this section, we present a conceptual mapping between explainability models and the behavior explanation categories defined by our framework. This mapping is essential both for selecting the appropriate model to answer a user's question and for excluding techniques unlikely to foster user trust.

In addition to the tools provided by Alibi and Qdrant, we also propose several advanced models not yet integrated into our stack. These require additional implementation effort but may offer substantial value for specific use cases.

\subsection{Explanation Models Mapped to Malle's Framework Categories}\label{subsec:preprocessing}
\begin{itemize}
    \item \textbf{Counterfactual Explanations}: Counterfactual explanations project alternative scenarios, simulating "what-if" conditions by tweaking input variables to achieve a desired outcome. \textbf{Available in Alibi.}

    \item \textbf{Prototypes}: Prototypes provide examples that best represent a particular class or decision boundary, forming a mental structure for users. \textbf{Available in Alibi under prototype-based explanations.}

    \item \textbf{Influential Instances}: Influential instances highlight specific training data points that most directly affect the model’s predictions, enabling a recall of their impact. \textbf{Available in Alibi.}

    \item \textbf{Global Surrogate Models}: Surrogate models approximate the original model’s behavior, forming a simplified structure for understanding global patterns. \textbf{Available in Alibi.}

    \item \textbf{Prototypical Examples Search}: Use the vector database to find the most representative examples (prototypes) of each class based on vector similarity. \textbf{Will be implemented in Qdrant.}

    \item \textbf{Criticism Detection}: Identify outlier instances (criticisms) that differ significantly from the core cluster of data points in the embedding space. \textbf{Will be implemented in Qdrant.}

    \item \textbf{Counterfactual Search}: Query the database for points closest to the decision boundary but belonging to a different class, enabling counterfactual reasoning. \textbf{Will be implemented in Qdrant.}

    \item \textbf{Influential Instances Retrieval}: Retrieve training examples that most influenced specific model predictions using similarity or validation metrics stored in the vector DB. \textbf{Will be implemented in Qdrant.}

    \item \textbf{Cluster Analysis}: Perform clustering on embedding vectors to identify groups of similar instances and their underlying characteristics. \textbf{Will be implemented in Qdrant.}

    \item \textbf{K-Nearest Neighbors (KNN) Classification or Regression}: Use a KNN model trained on the embeddings to classify or predict based on neighboring examples in the database. \textbf{Will be implemented in Qdrant.}

    \item \textbf{Prototype-Based Surrogate Models}: Build surrogate models that rely on the most representative prototypes for each class or decision boundary. \textbf{Will be implemented in Qdrant.}

    \item \textbf{Feature Attribution Models}: Combine vector db embeddings with feature attribution techniques like SHAP or LIME to explain the relative importance of features for each prediction. \textbf{Will be implemented in Qdrant.}

    \item \textbf{Covariation Analysis for Model Validity}: Use stored labels and validation results to identify patterns of covariation between model errors and input features or clusters. \textbf{Will be implemented in Qdrant.}

    \item \textbf{Adversarial Example Detection}: Identify instances near the decision boundary that are most sensitive to perturbations, using vector db embeddings for boundary searches. \textbf{Will be implemented in Qdrant.}

    \item \textbf{Causal Validation Models}: Train models that explicitly test causal relationships between features and labels based on validated results. \textbf{Will be implemented in Qdrant.}
    \item \textbf{Case-Based Reasoning (CBR)}: Leverages previous cases similar to the current instance to inform and explain model predictions, helping users relate decisions to familiar contexts.

    \item \textbf{Exemplar-Based Explanations}: Provides representative examples (exemplars) from the dataset that most closely resemble the current input, facilitating understanding through direct recall of past cases.

    \item \textbf{Memory-Augmented Neural Networks (MANNs)}: Enhances neural networks with external memory modules to retrieve relevant historical data, grounding explanations in previously encountered scenarios.

    \item \textbf{Nearest Prototype Recall}: Identifies and presents the prototype (typical example) from a learned set of data points that best matches the current input, enabling intuitive comparisons.

    \item \textbf{Historical Decision Logs}: Uses recorded logs of past model decisions and inputs to offer concrete, historical context for current model outputs, aiding transparency and recall.

    \item \textbf{Natural Language Explanations (NLE)}: Converts complex model decisions into human-understandable language, providing accessible, narrative-like rationalizations of predictions.

    \item \textbf{Decision Rule Extraction (RuleFit)}: Extracts interpretable rules from complex models to rationalize predictions in clear, rule-based formats.

    \item \textbf{Narrative-Based Explanations}: Constructs stories or sequential accounts to contextualize model behavior, making explanations more relatable and cognitively engaging for users.

    \item \textbf{Post-Hoc Justifications (Bias Alignment)}: Provides justifications after the decision-making process, aligning explanations with user expectations or ethical considerations to foster trust.

\end{itemize}

\begin{table*}[ht]
\centering
\renewcommand{\arraystretch}{1.2}
\setlength{\tabcolsep}{4pt} 
\footnotesize 
\begin{tabular}{|l|c|c|l|}
\hline
\textbf{Explanation Model} & \textbf{Alibi Availability} & \textbf{VectorDB Implementation} & \textbf{Malle's Category} \\
\hline
\textbf{Counterfactual Explanations}         & \ding{51} & \ding{55} & Simulation/Projection \\
\textbf{Adversarial Examples}                & \ding{55} & \ding{51} & Simulation/Projection \\
\textbf{Prototypes}                          & \ding{55} & \ding{51} & Knowledge Structures \\
\textbf{Criticisms}                          & \ding{55} & \ding{51} & Knowledge Structures \\
\textbf{Influential Instances}               & \ding{55} & \ding{51} & Direct Recall \\
\textbf{K-Nearest Neighbors (SHAP-based)}    & \ding{55} & \ding{51} & Knowledge Structures \\
\textbf{Partial Dependence Plot (PDP)}       & \ding{51} & \ding{55} & Covariation \\
\textbf{Accumulated Local Effects (ALE)}     & \ding{51} & \ding{55} & Covariation \\
\textbf{Feature Interaction (H-statistic)}   & \ding{51} & \ding{55} & Covariation \\
\textbf{Functional Decomposition}            & \ding{51} & \ding{55} & Knowledge Structures \\
\textbf{Permutation Feature Importance}      & \ding{51} & \ding{55} & Covariation \\
\textbf{Global Surrogate Models}             & \ding{51} & \ding{55} & Knowledge Structures \\
\textbf{Model Training Report}               & \ding{55} & \ding{51} & Rationalization \\
\textbf{Cluster Analysis for Covariation}    & \ding{55} & \ding{51} & Covariation \\
\hline
\textbf{Case-Based Reasoning (CBR)}          & \ding{55} & \ding{55} & Direct Recall \\
\textbf{Exemplar-Based Explanations}         & \ding{55} & \ding{55} & Direct Recall \\
\textbf{Memory-Augmented Neural Networks (MANNs)} & \ding{55} & \ding{55} & Direct Recall \\
\textbf{Nearest Prototype Recall}            & \ding{55} & \ding{55} & Direct Recall \\
\textbf{Historical Decision Logs}            & \ding{55} & \ding{55} & Direct Recall \\
\hline
\textbf{Natural Language Explanations (NLE)} & \ding{55} & \ding{55} & Rationalization \\
\textbf{Decision Rule Extraction (RuleFit)}  & \ding{55} & \ding{55} & Rationalization \\
\textbf{Narrative-Based Explanations}        & \ding{55} & \ding{55} & Rationalization \\
\textbf{Post-Hoc Justifications (Bias Alignment)} & \ding{55} & \ding{55} & Rationalization \\
\hline
\end{tabular}
\caption{Explanation Categories and Their Implementations}
\label{tab:explanation_methods}
\end{table*}

Now we have a series of ingredients that, when properly selected, can be used to build an explainability framework. As with all ingredients, some are equivalent, while others are complementary.

Using our framework, we are able to clearly categorize any system as suited for either \textit{Action}-based or \textit{Experience}-based explanations. However, selecting a single implementation method remains challenging due to dependencies on external or context-specific factors that cannot be fully addressed during the model design phase.

In data-driven products, such challenges are typically managed through recommendation systems calibrated on user preferences. Similarly, in our context, users should initially be allowed to choose the methods they are most comfortable with. Over time, an adaptive system can be introduced to guide them toward the most effective or widely used methods. Until such a system is in place, we rely on the general heuristics proposed by Malle~\cite{malle2004}.

The framework serves not only to select the most appropriate methods for explainability, but also to discard those that are misaligned with the actual use case. In the following two sections, we present practical case studies where the framework is applied to assess the coherence of commonly used explainability techniques and demonstrate how their refinement naturally emerges through structured application.

\subsection{Application to Credit Scoring}\label{subsec:crd}
We apply our explainability framework to a credit scoring system (similar to BUSSMANN~\cite{bussmann2020}) used in corporate loan approvals. The system processes financial data to assign a creditworthiness rating, determining loan eligibility based on a predefined threshold. If a rating falls below this threshold, loan officers may propose an override—provided they justify their decision.

This scenario exemplifies an AI-Assisted Decision-Making (ADM~\cite{bussmann2020}) system, where understanding model-driven rejections is crucial for agents seeking to refine their decision strategies. Typically, agents use the tool in loan approval processes they initially expect to yield positive outcomes. Any rejection by the model is therefore perceived as an intentional obstruction of their task and requires a clear explanation. This expectation positions the model’s role as one of \emph{emulation}—mirroring the agent’s judgment while integrating additional qualitative insights. In terms of our framework, this corresponds to the \emph{Actions} case:

\begin{center}
\textit{Emulation} $\sim$ \textit{Observable} $\rightarrow$ \textit{Intentional} $\sim$ \textit{Actions}
\end{center}

Moreover, it is important to note that whenever we seek explanations for negative outcomes, we are operating in the domain of intentional behavior~\cite{malle2004}. When the model's outputs align with the agent’s expectations, explanations are typically not requested—the agent assumes the system is functioning properly. As a result, explainability demands are concentrated around adverse outcomes, reinforcing our categorization.

We will consequently rely on the following explanation types:

\begin{itemize}
    \item \textbf{Reason Explanations (REA)}: Link decisions to specific rationales (e.g., identifying risky financial traits).
    \item \textbf{Causal History of Reasoning (CHR)}: Attribute decisions to broader statistical patterns.
    \item \textbf{Enabling Factors (EF)}: Highlight external conditions (e.g., macroeconomic trends) influencing the outcome.
\end{itemize}

Considering the specific characteristics of this task—including judged behavioral attributes, pragmatic goals, and available information resources, as outlined by Malle~\cite{malle2004}—we can disregard Enabling Factors and rely solely on the first two explanation strategies. The Enabling Factor is of minor importance in this context and may not be convincing to the agents, who are likely to dismiss such explanations as ill-posed or irrelevant.

\paragraph{1. Causal History of Reasoning (CHR): Trends Over Time}
When rejections align with broader data trends, CHR explanations are appropriate and as stated in Table~\ref{tab:explanation_modes} are often lined to Malle’s category of Knowledge Structures. Effective knowledge structure techniques include:
\begin{itemize}
\item \textbf{Prototypes}, \textbf{Criticisms}, \textbf{Functional Decomposition}, and \textbf{Global Surrogate Models}.
\item \textbf{K-Nearest Neighbors (KNN)}: Particularly intuitive for agents, offering transparency through similarity to past observed cases.
\end{itemize}

\paragraph{2. Reason Explanations (REA): Singular Decisions}
When decisions appear isolated or context-specific, REA explanations—aligned with Malle’s category of Direct Recall—are more suitable. Recommended methods include:
\begin{itemize}
\item \textbf{Influential Instances}: Highlight key features that directly influenced the decision.
\item \textbf{Similar Historical Data}: Provide contextual comparison with analogous decisions made in the past.
\end{itemize}

We have therefore identified Knowledge Structures and Direct Recall as the most suitable categories for our explainability approach, deliberately excluding alternative methods. The next step is to select the appropriate explainability model to implement.

The specific choice of mathematical implementation depends on the type of model employed and the availability of data. In our system, we compute ratings following the methodology proposed in~\cite{provenzano2020mlcreditscoring}. Operating within the domain of balance sheet data, we benefit from high data availability and access to historical evaluations. Given this context, we adopt a K-Nearest Neighbors approach (augmented with SHAP values) and serve it through a vector database. This setup enables the implementation of Prototypes and Criticisms, allowing us to highlight both similar and contrasting examples relative to the evaluated case.

In addition, we compute Feature Importance within size-sector quadrants, focusing on Influential Instances, and provide visual charts based on validated data to aid interpretation. This design ensures that our explainability tools remain aligned with the selected categories—Knowledge Structures and Direct Recall—thus enhancing user comprehension and trust.

\begin{figure}[h]
    \centering
    \includegraphics[width=0.8\linewidth]{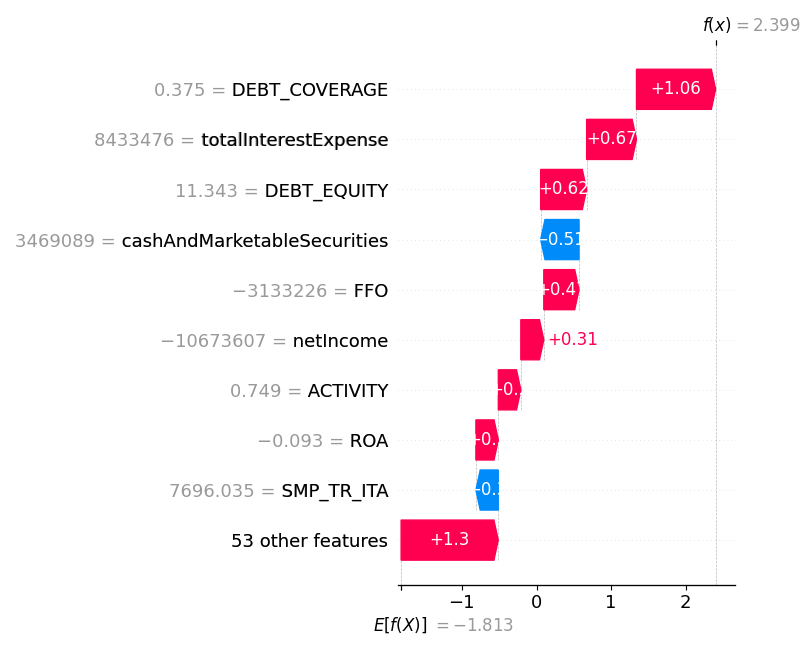}
    \caption{Shap}
    \label{fig:shap_example}
\end{figure}

As previously mentioned, our framework can also be used to exclude explainability methods that are unlikely to foster trust. A commonly adopted approach is the use of pure \textbf{SHAP values}, which decompose an individual prediction into additive feature contributions, as shown in Figure~\ref{fig:shap_example}. While informative, this technique—when applied to a single instance—primarily corresponds to the Simulation/Projection category, where the user mentally simulates how changes in inputs affect the output. This kind of reasoning may be relevant for Causal History explanations, but only after Knowledge Structures-based approaches have been exhausted.

To summarize, effective explainability for ADM credit scoring models hinges on presenting users with relevant examples through Causal History and Direct Recall-aligned methods. All other techniques are either less effective or inappropriate in this context.

\subsection{Application to Documentation Analysis}\label{subsec:crd1}

Explainability in Automated Decision-Making (ADM) has gained increasing importance with the rise of generative language models. While traditional supervised models offered a certain degree of explainability, generative Large Language Models (LLMs) operate in social and interactive contexts where trust is critical. In classification or regression tasks, users can anticipate occasional prediction errors. However, generative models introduce the additional risk of hallucinations—confident but incorrect outputs—which can severely undermine trust in the system.
For these systems, explainability is deeply intertwined with trust, which can only be achieved through proper behavioral explanations.

In many cases, no explicit explanation is provided—only a generated, often lengthy argument crafted through prompt engineering. In more advanced systems, the model may highlight key documents or sentences that influenced its response. But is this approach effective? And more importantly, is it enough to ensure transparency and trust?

\textbf{Framework Overview:} For documentation analysis, the explainability pattern follows:

\begin{center}
  \textit{Discovery} $\sim$ \textit{Unobservable} $\rightarrow$ \textit{Unintentional} $\sim$ \textit{Experiences}
\end{center}

This reflects the exploratory nature of document analysis, where answers are not directly measurable, and reasoning may be implicit or unintentional.

In this case, explanations fall under Unintentional Causes:
\begin{itemize}
    \item \textbf{Person Causes}: Attributing responses to the model’s inherent characteristics or training data.
    \item \textbf{Situation Causes}: Linking outputs to document context or retrieval mechanisms.
    \item \textbf{Various Interactions}: Explaining how model structure and document context influence generated responses.
\end{itemize}
Since document analysis lacks a human decision-maker, the primary explanatory mode falls under Situation Causes, where the focus is on the external source material influencing the system's response. According to Table~\ref{tab:explanation_modes}, this places us within the Reason/Actors category, best represented by the Direct Recall method.

The most common Direct Recall model applied to LLM-based systems is Influential Instances. By highlighting key passages from reference documents, this method reveals the specific knowledge the LLM relied on. Rather than offering causal reasoning or hypothetical alternatives, it traces the model’s outputs back to their sources. A practical implementation involves using a vector database that stores document embeddings, functioning similarly to k-nearest neighbors in a semantic space. This allows the system to retrieve the most relevant passages based on their similarity to the user's query.

To further enhance trust while remaining within the Direct Recall paradigm, we propose extending from Situation Causes to Person Causes. Introducing human-generated data adds a social interaction dimension to the model. The implementation of Historical Decision Logs represents a promising approach to this extension. This involves incorporating previously validated responses into the vector database. When a user query matches a reviewed document, the system can provide the past answer, shifting explainability and trust toward the original human agent who validated the response.

Our framework, when applied to rejection scenarios, indicates that other common LLM explanation methods—such as providing confidence intervals or probabilities of answer correctness—should be used cautiously, or even avoided in certain contexts. These techniques belong to the Simulation/Projection category and are typically appropriate for explaining Intentional behaviors, which Malle’s framework classifies as Actions. However, when applied to systems that primarily generate Experiences—outputs without a clear ground truth—these methods may yield unintended consequences.

Specifically, the inclusion of confidence-based metrics implicitly frames the model as producing accurate or expected results. This shifts the user's mental model from viewing the system as exploratory to treating it as deterministic and evaluable. Such a shift—from Experiences to Actions—can backfire: rather than increasing user trust through transparency, it may diminish trust by creating false expectations of certainty and control, especially when the system fails to meet them.

In summary, while confidence-based metrics might appear to offer clarity, they risk undermining trust in Experience-oriented systems. A sociological approach based on Direct Recall models avoids this pitfall by aligning the explanation type with the user’s interpretive frame.

\section{Conclusions}\label{sec:conclusions}

In this paper, we explored the distinction between emulation and discovery tasks in the context of explainable AI, establishing a conceptual bridge between Malle’s behavior explanation framework and the technical paradigms of supervised and unsupervised learning. By aligning sociological categories with machine learning structures, we developed a principled method for selecting appropriate explanation tools based on the type of behavior being modeled—whether intentional (Actions) or unintentional (Experiences).

Through a structured equivalence between explanation techniques and behavioral categories, we were able not only to recommend suitable methods but, crucially, to identify and exclude those that are likely to reduce user trust when misapplied. This enables product managers and modelers to avoid the common pitfall of applying technically valid but contextually inappropriate tools. The two case studies presented demonstrated how the framework can be applied in practice, highlighting the risks of widely used—but often unsuitable—explanatory strategies.

Looking ahead, future research will focus on refining the personalization of explainability by tailoring explanations to the user's level of expertise, contextual needs, and cognitive preferences. Achieving this goal will require the integration of generative language models capable of dynamically producing diverse explanatory formats—such as marked versus unmarked beliefs, or traits versus non-traits—rooted in cognitive science. This adaptive, user-centered approach holds the potential to deliver explanations that are not only technically accurate but also psychologically resonant, thereby enhancing both trust and usability in AI-assisted decision-making systems.

\section*{Acknowledgments}

We would like to extend our gratitude to Claudio Nordio for his role as Chief Risk Officer of illimity Bank, which supported the overall progress of this work.

\bibliographystyle{plain}
\bibliography{biblio}

\begin{thebibliography}{10}

\bibitem{ms_responsible_ai}
Microsoft AI.
\newblock Responsible ai: Overview and best practices, 2024.

\bibitem{bussmann2020}
Niklas Bussmann, Paolo Giudici, Dimitri Marinelli, and Jochen Papenbrock.
\newblock Explainable machine learning in credit risk management.
\newblock {\em Computational Management Science}, 17(4):885--905, 2020.

\bibitem{chen2022}
Chacha Chen, Shi Feng, Amit Sharma, and Chenhao Tan.
\newblock Machine explanations and human understanding.
\newblock {\em arXiv preprint}, 2022.

\bibitem{ibm_ai}
IBM~Design for AI.
\newblock Ai conversation planning, 2024.

\bibitem{guidotti2018_arxiv}
Riccardo Guidotti, Anna Monreale, Salvatore Ruggieri, Franco Turini, Fosca
  Giannotti, and Dino Pedreschi.
\newblock Explainable artificial intelligence: A systematic review.
\newblock {\em arXiv preprint}, 2018.

\bibitem{guidotti2018}
Riccardo Guidotti, Anna Monreale, Salvatore Ruggieri, Franco Turini, Fosca
  Giannotti, and Dino Pedreschi.
\newblock A survey of methods for explaining black box models.
\newblock {\em ACM Computing Surveys}, 51(5):93, 2018.

\bibitem{gift2024}
Gift Living.
\newblock Explainable ai in credit scoring models.
\newblock {\em ResearchGate Preprint}, 2024.

\bibitem{malle2004}
Bertram~F. Malle.
\newblock {\em How the Mind Explains Behavior: Folk Explanations, Meaning, and
  Social Interaction}.
\newblock MIT Press, 2004.

\bibitem{google_explainability}
Google PAIR.
\newblock Explainability and trust in ai, 2024.

\bibitem{pelkmann2024}
Matthias Pelkmann and Stefan Feuerriegel.
\newblock Explainable ai in finance: A systematic literature review.
\newblock {\em Artificial Intelligence Review}, 2024.

\bibitem{provenzano2020mlcreditscoring}
A.~R. Provenzano, D.~Trifirò, A.~Datteo, L.~Giada, N.~Jean, A.~Riciputi,
  G.~Le~Pera, M.~Spadaccino, L.~Massaron, and C.~Nordio.
\newblock Machine learning approach for credit scoring.
\newblock {\em arXiv preprint}, 2020.
\newblock arXiv identifier or journal details missing.

\bibitem{samek2019}
Wojciech Samek, Grégoire Montavon, Andrea Vedaldi, Lars~Kai Hansen, and
  Klaus-Robert Müller.
\newblock {\em Explainable AI: Interpreting, Explaining and Visualizing Deep
  Learning}.
\newblock Springer Nature, 2019.

\bibitem{schmitt2024}
Marc Schmitt.
\newblock Explainable automated machine learning for credit decisions:
  Enhancing human artificial intelligence collaboration in financial
  engineering.
\newblock {\em arXiv preprint}, 2024.

\bibitem{qdrant}
Qdrant Team.
\newblock Qdrant: High-performance vector search engine, 2024.

\bibitem{alibi}
Seldon Technologies.
\newblock Alibi: Algorithms for explainability, 2024.

\bibitem{tyagi2022}
Swati Tyagi.
\newblock Analyzing machine learning models for credit scoring with explainable
  ai and optimizing investment decisions.
\newblock {\em arXiv preprint}, 2022.

\bibitem{zhang2020}
Yunfeng Zhang, Q.~Vera Liao, and Rachel K.~E. Bellamy.
\newblock Effect of confidence and explanation on trust calibration in
  ai-assisted decision making.
\newblock {\em arXiv preprint}, 2020.

\end{thebibliography}

\end{document}